# Quantifying the loading capacity of a carrier-based DPI formulation and its dependence on the blending process

Andrea Benassi[1*], Ilaria Perazzi[2], Roberto Bosi[1], Ciro Cottini[1], Ruggero Bettini[2]

1-DP Manufacturing & Innovation, Chiesi Farmaceutici SpA, Parma Italy

2-Department of Food and Drugs, University of Parma, Parma, Italy

[*]Corresponding author address: Chiesi Farmaceutici S.p.A. Largo Belloli 11A– *43122 Parma (Italy)*

**Abstract**

Non-segregating ordered powder mixtures constituted by a coarse carrier fraction and finer components are at the basis of *dry powders for inhalation* pharmaceuticas. The estimation of the loading capacity, i.e. how many fines can be hosted on each carrier particle, is crucial to grant the product quality through a reproducible and affordable manufacturing process. We propose an approach based on the combination of sieve analysis, optical microscopy and powder bed permeability to quantify the loading capacity and understand the fines behavior, the impact of the mixing process was also investigated. We tested the method on model binary mixtures composed only of a coarse lactose carrier and micronized lactose fines as a surrogate of a real active pharmaceutical ingredient. The results provided by the different methods are consistent, the approach proved to be accurate and reproducible. The effect of different mixing parameters and equipment on the loading capacity is also discussed.

## 1. Introduction

Dry powders for inhalation have gained an increasing deal of interest over the last decades for the undoutable adavantages offering compared to other method of drug administration to the lung. Adhesive powder mixtures constituted by a micronized active pharmaceutical ingredient (API) and a coarse carrier such as lactose are the mosly represented especially for the treatment of deseases such as COPD or asthma. The effect of the interaction between the powder components on the aerosolization performance, and thus on the therapy efficacy, both from the physico-chemical and mechanical point of view is far from being fully understood, as well as the effect of the manufacturing process parameters [1]. Several aspects such as carrier physico-chemical properties [2] surface topography [3], solid state [4], electrostatic charging [5] combine and interplay in determining the dispersibility of micronized API in these formulations. Besides the aerosolization performance has been reported to be influenced by formulation parameters such as the dose of API [6–9], the presence of intrinsic or extrising fines [10,11] or process-related factore such as blending procedure [12,13] or mixing time [9].

Among the formulation-related aspects, the drug-carrier ratio and the loading capacity of the latter are gaining a great deal of interest since the therapeutic doses carried by these formulation vary significantly ranging from 0.1% to more thena 4% by weight [9].

Taking into consideration this latter point the API's physico-chemical and solid-state properties further complicates the picture, especially if one considers that the size reduction is obtained by processes such as jet milling or spray drying that may significantly impact on said properties. The complexity of real carrier-based formulations, composed by several different excipient and ternary agents for lubrication or passivation of too active bonding sites [14], may be dropped in favour of a simplified picture catching only the basic physics and features of such pharmaceutical form. Following this approach Rudén et al. [15] recently investigated the relationship between the mixture performance and the surface coverage ratio using micronized lactose as a surrogate of the API. They concluded that there is still a need of investigation based on physical properties to further elucidate the relationship.

In the present paper we investigated the suitability of three different physical methods, i.e. sieve analysis, optical microscopy and permeability, to quantify the loading capacity of a lactose carrier using micronized lactose as substitute of API in adhesive mixture for inhalation. A further goal was to assess whether mixing process characterized by different energetics might affect the loading capacity and the aerosolizatioin performance.

Althoug the first adopted method is well established in the pharmaceutical technology fied, it is seldom used for quantification other than particle size distribution. On the other hand powder rheology test such as the permeability measurement, are acquiring increasing importance in powder technology area [16–18].

Here we propose an approach based on the combination of different methods to contribute at elucidating the effect of the saturation of the carrier surface by the added fines focusing mainly on formulation and process perspectives rather than on the aerosolization performaces, though the latter was considered mainly as a mean to assess the blend uniformity and the effect of different mixing processes on the carrier loading capacity.

## 2. Materials and Methods

**2.1 Materials and blend composition**

CapsuLac®60 (Meggle) represents a rough and coarse α–monohydrate lactose carrier which has been sieved retaining only particles with diameter in the range 212-355 µm, more specifically dv10 = 198.5 µm , dv50 = 300.5 µm and dv90 = 440.3 µm. LactoHale® 300 (DFE Pharma) mimics an active pharmaceutical ingredient with dv10 = 0.4 µm , dv50 = 2.7 µm and dv90 = 6.7 µm. SEM images and particle size distributions for such excipients are shown in Figure 1, the particle size distribution of the two excipients was obtained with a Helos BR from Sympatec® using 0.5 bar with an R5 lens for Capsulac® 60 and 2.0 bar with an R1 lens for Lactohale® 300. Different blends have been prepared with a concentration ranging from 1 to 30% w/w of fine lactose into coarse carrier.

**2.2 DPI Blending**

Dry powders for inhalation (DPI) blends have been prepared with both a low shear tumbling mixing process, through a Turbula® T2F (WAB AG, Switzerland), and a high shear process, through a Cyclomix® IM-5 (Hosokawa Micron B.V., Japan). In both cases, as a starting condition, the fine lactose mimicking the API has been sandwiched between two layers of coarse carrier of equal thickness. Before blending the aliquotes of LH300 fines did not contain any significant amount of aggregates appearing perfectly uniform both at visual inspection and to a closer look with the optical microscope. To ensure the complete absence of any pre-existing aggregate one might be tempted to sieve the LH300 fines before blending. However sieving a fine powder such as LH300 (or any real inhalable API) could be detrimental to its homogeneity: the strong cohesion between particles prevent them from flowing through the sieve spontaneously, a mechanical action is necessary to force them to pass through the net, resulting in powder compaction and tribo-charging eventually leading to agglomeration.

For the low shear mixing process a 1.5 L bin has been employed; in order to estimate the impact of the process parameters on the loading capacity, different blends were pepared with batch size 250 and 500 g (to evaluate the effect of the different filling degree of the bin), with mixing speed from 23 to 72 revolutions per minute (rpm) and with mixing time from 7 to 180 minutes.

A bin with an internal geometry specifically designed to increase the shear planes was also employed, hereafter referred as "middle shear" mixer.

High shear blends were prepared with batch size 1 Kg for a total mixing time ranging in the interval 5-20 minutes at 250 rpm. In these mixing conditions a relatively high energy was transferred to the powder in a short time without damaging the larger carrier particles and without resulting in comminution and size reduction effects.

**2.3 Sieve analysis**

To separate the non-bonded fines from the blended carrier a sieve shaker AS200 (Retsch, Germany) was employed mounting two sieves selected according to the carrier particle size distribution. As illustrated in Figure 2 (a), a 425 $\mu$m sieve retained all the fine aggregates letting the blend pass below, a 180 $\mu$m sieve stopped all the blend particles letting only the free fines or very small powder aggregates to pass below end being collected in the lower dish. The adopted sieving protocol is based on the idea of transferring very little energy to the powder in order to avoid fine detachment from the carrier and shear forces between particles. This was achieved by sieving a limited amount of powder, 50 to 80 g at 50 Hz and with a small oscillation amplitude of 0.2 mm for 15 minutes. We first verified that sieving a uniform blend at 10% w/w fine concentration in these conditions did not result in fine detachment, this was accomplished by monitoring in real time the weight of the lower collection dish containing the free fines: this quantity of powder did not increase with increasing sieving time up to 15 minutes. We then verified that, in case of high fines concentration (30% w/w), where fine aggregates did form and were stopped by the 425 μm sieve, their exposition to mechanical vibrations did not alter significantly their size and shape in time. If this was the case de-aggregated fines could have been released (by abrasion and inter-aggregate collisions) in the uniform blend underneath compromising the measurement of the real state of the fines after mixing. After 15 minutes the net weight of each sieve was measured and the percentage over the total initial mass was calculated for the fine aggregates stopped in the upper sieve and the non-bonded fines found in the collection dish. The amount of fines bonded to the carrier was then estimated by subtracting these quantities to the nominal fine content.

The reproducibility of the method has been verified as well: data reported in the present work represent the average of at least 3 distinct sieving tests executed on different aliquots of the same blend. The relative standard deviation is displayed through the error bars in the plots and it is almost always smaller than 1%. For small and large fine concentrations blending has been repeated to test also the batch-to-batch variability, the relative standard deviation is smaller than 1%, i.e. comparable to the intra-batch standard deviation.

It must be stressed that LactoHale® aggregates with size comparable to that of the carrier particle, and more generally smaller than 425 μm, might exist and they cannot be separated from the rest of the blend. Although the sieved blend may look homogeneous to a visual inspection, such small aggregates could be easily put into evidence with an optical microscope, see Figure 2 (b).

Therefore, it is clear that the fines trapped in these non-sieveble aggregates counts as "bonded fines" slightly altering the correct estimation. However, the amount of these aggregates was generally small and a way to minimize their presence is discussed later on.

Finally, the large aggregates stopped by the 425 μm sieve are shown in Figure 2 (c) and (d). By breaking them under the optical microscope it was easy to verify that they were made entirely by LactoHale® 300 and only rarely they incorporated a coarse carrier particle, likely working as a seed for the growth of the thick LactoHale® coating layer. The appearance of a broken blend particle and a broken aggregate is shown in Figure 2 (e) and (f) respectively.

**2.4 Permeability measurement**

The measure of the permeability is sensitive to the amount of fines present in the blend as they fill the pores and cavities of the irregular carrier particles as well as the interstitial regions between them. This determines a size reduction of the channels available to the air to percolate through the powder with a consequent increase of speed and drop of pressure. Moreover, the presence of fine patches on the otherwise more smooth carrier surface increases locally the surface roughness, thus increasing both the viscous and turbulent dissipation at the channels boundaries according to the Darcy- Weisbach theory [19].

This techniques is widely employed in the literature both on pure excipients, to assess the particle properties [20], and on binary or ternary mixtures e.g. to predict the DPI performances [21].

An aliquot of each sieved blend was used to measure the permeability $\kappa$ through the powder rheometer FT4 (Freeman Technology, UK). The instrument measures the pressure drop $\Delta p$ of an air flow, with dynamic viscosity $\mu$ volumetric flow rate $Q$ and average speed $v$, as it passes through a cylindrical cuvette of radius $R$ filled with the blend powder. A vertical piston exerts a pressure on the powder compacting it and measuring the total powder bed height $L$. The permeability $\kappa$ is then calculated via Darcy law [22]:

$$\kappa = \frac{Q\,\mu\,L}{\pi R^2\,\Delta p} = \frac{v\,\mu\,L}{\Delta p} \qquad (1)$$

The permeability measurements were carried out after gently conditioning the blends with a rotating blade to remove the inhomogeneities due to pouring and after compressing the powder to 15 kPa with a vertical piston [23]. In the measurements here presented a cylindrical glass vessel with radius *R = 25 mm* has been employed to host the powder bed, corresponding to a powder volume of 10 mL, the air flow rate has been set to *v= 2mm/s*. The presented data are obtained as the average over at least three tests, the standard deviation is represented through the error bars.

**2.5 Carrier coverage analysis through optical microscopy**

Optical micrographies of Figure 3 (a) clearly evidence that the coverage of the carrier surface, represented by the white patches of fines, increases with increasing the LactoHale® 300 concentration in the blend. This is true at least for the first pictures from the left hand side while, at higher concentrations, it is hard to see any further increase of the coverage. To substantiate with numbers this visual sensation we exploited the clear difference in brightness among the black background, the carrier surface and the fine patches. Filtering the pixels of each micrography by their brightness we separated and counted the number of pixels (and thus the total area) occupied by the powder particles and the subset of pixel representing the patches of fine. An example is given in Figure 3 (b) where the overall particle area is marked in red and the patches of fine are marked in green. A dimensionless coverage index was defined as the ratio between the patches of fine and the total particle areas, it ranged between 0 and 1, approaching 1 if the carrier was completely covered. The coverage index for the blends of Figure 3 (a) is given in panel (c) as a function of the concentration of fines; the relationship was liner up to about 25% w/w of fine in the blend, thereafter it tended to level off.

The images were acquired with a AZ100M microscope (Nikon, Japan) and their filtering performed with the native software NIS-Elements (Nikon, Japan) . The brightness of each pixel is represented by an integer number in the range 0-255 (black-whithe), for every micrography a brightness histogram can be build binning the pixels according to this number, an example in given in the bottom part of Figure 3 (b), the spike on the left side represents the black background pixels. We have set every pixel with brightness > 100 as pertaining to the full blend particles, this fist threshold corresponds to the first minimum observed in the histogram moving to the right (red dashed line). We then set all the pixels with brightness > 190 as pertaining to the fine patches, this corresponds to the second maximum found on the right side of the histogram (green dashed line). These thresholds are clearly dependent on the lightening of the samples and on the optical properties of materials employed for the blends, thus the comparison of coverage indexes taken on the same materials with the same instrument and in the same lightening conditions, makes perfect sense while comparing coverage indexes measure with different instruments and on different materials is not so straightforward.

The coverage indexes have been calculated averaging over 6 pictures containing 30 to 50 particles each one for a total of 250-300 particles.

Finally it must be stressed that the coverage index was not directly proportional the amount of LactoHale® 300 fines carried by the Capsulac® 60 particles. The coverage analysis was in fact performed on the two-dimensional projection of the particle images thus it was not able to account for the thickness of the fine patches or their density. An amount of LactoHale® 300 could also be trapped into pores and interstitial regions remaining underneath the visible patch.

The arbitrariness in the choice of the optical intensity threshold and the limitations of working with two-dimensional projections of the particles above mentioned are also present in chemical imaging techniques. Raman, NIR or x-ray-based topography images, typically associated with SEM images of the same sample, always appear as colored patches whose blurred boundary fade over a dark noisy background, thus an arbitrary threshold to define whether a boundary pixel pertain to the fine patch or the background must still be introduced. On the contrary simple optical microscopy has three main advantages:

1- It can be used to study binary mixtures featuring a single material, like our lactose-on-lactose case;

2- It is faster as it does not require to collect spectroscopic information about the different materials and, e.g., to calibrate detectors on a specific emission peak;

3- With large field images one can collect data relevant to a lot of particles with a limited number of pictures and samples. Most of the chemical imaging techniques have a very limited optical field requiring particles to be imaged one by one, a lot of effort might be necessary to collect a statistically representative pool of data.

**2.6 Aerosolization performances and blend uniformity assessment**

Several literature papers deal with the relationship between the loading capacity and the aeroslization performance of different carries [6,7] aiming at demonstrationg the effect of the percentage of loaded drug on the *in vitro* respirability. Here, a slightly different approach was adopted: the aerosolization performances was first of all investigated as a tool to evaluate the uniformity of distribution of fines, thus to pursue the main scope of the project that was the evaluation of new procedure for investigating the loading capacity and the effect of the mixing process. As a matter of fact the prepared blends were constituted solely by lactose, therefore, there was no chance to apply an assay for the fine component and its standard deviation across the mixing bin with HPLC or spectroscopic methods. Indeed should certain regions of a blend be very rich of fines compared to other portions of the same batch, differences in the fine fraction measured for capsules sampled in those blend regions are expected to be observed. The inhalation performances presented later in this work are obtained by manually filling capsules (HPMC Vcaps, size 3 capsules Capsugel ®, Switzerland) with 20 ± 1 mg of blend sampled randomly from different regions of the mixing bin. The capsules have been subsequently discharged with an RS01 single capsule device (RPC Plastiape®, Italy) through a Fast Screening Impactor (Copley Scientific®, UK) operated at a flow rate of 60 L/min for 4 sec. The fine particle mass has been measured by weighing the powder deposited in the latest, and finest (equivalent diameter < 5 μm), filter separately for each capsule, average and standard deviation are calculated over the 5 capsule discharges. We considered the standard deviation over the capsule performances representative of the uniformity of each blend batch. By weighing the device with the capsule before and after the discharge the residue of powder left inside can be estimated and thus the emitted dose was calculated as the difference from the loaded dose and said residue. Finally the fine particle fraction (FPF) was calculated normalizing the fine particle mass by the emitted dose.

## 3. Results and Discussion

Before focusing on the blend behaviour as a function of increasing fine concentration and on the carrier loading capacity it was necessary to investigate deeply the blending process itself. This is in fact the means by which we generated the starting powder mixtures, we thus needed to understand and control the impact of the process parameters on the overall blend characteristics.

**3.1 Mixing as a steady state process**

Several blends at 27% w/w concentration have been prepared with both low and high shear processes stopping the mixing after fixed time intervals, from 5 to 180 minutes. The blends were sieved according to the procedure previously described and the weight of free, bonded and agglomerated fines were plotted against the mixing time. The obtained results are shown in Figure 4 (a) and (b) by the colored markers; a plateau for the different fractions is reached in both cases after less than ten minutes. For the high shear case it was not possible to proceed beyond 20 minutes as for longer mixing times the carrier itself underwent comminution due to the high amount of energy transferred to the powder particles. The fine aggregate mass achieved a maximum after few minutes of mixing when the amount of available free fines was still considerable. This non-monotonic behaviour is easily explained by considering that while bonding to the carrier by adhesion, the fines also have a pronounced tendency to form aggregates by cohesion, thus their number initially increases; once the abundance of free fines drops down, the mechanism of aggregate collisions and abrasion becomes as relevant as aggregation reducing size and number of aggregates down to a constant value (or zero in case of high shear).

The presence of a plateau, i.e. a steady state condition, confirms that the blends were non-segregating mixtures as one could expect given the large difference in size between coarse and fine particles and the high cohesion of the fine fraction due to its large specific surface area [24].

The plateau mirrors the dynamic equilibrium among the three populations involved in the mixing process [25]. This is depicted in panel (c) of Figure 4: aggregates can grow form free fines with a rate $k_{growth}$; fines can be made free by aggregate abrasion or breakage with a rate $k_{break}$; fines can attach or detach from the carrier surface with rates $k_{on}$ and $k_{off}$ respectively; fines can be exchanged between aggregates and covered carrier particles during collisions with a frequency $k_{collisions}$. Neglecting the latter term for the sake of simplicity a simple set of first order differential equations can be written to describe the variation of the concentration of the three populations as a function of time, exactly as one would do for first order chemical reactions

$$\frac{\partial C_b(t)}{\partial t} = k_{on} C_f(t)(C^{max} - C_b(t)) - k_{off} C_b(t) \quad (2)$$

$$\frac{\partial C_a(t)}{\partial t} = k_{growth} C_f(t) - k_{break} C_a(t) \quad (3)$$

$$\frac{\partial C_a(t)}{\partial t} + \frac{\partial C_b(t)}{\partial t} + \frac{\partial C_f(t)}{\partial t} = 0 \quad (4)$$

Where $C_b(t)$, $C_a(t)$ and $C_f(t)$ are the concentrations of bonded fines, fine aggregates and free fines respectively; $C^{max}$ is the maximum concentration of fines allocatable on the carrier surface, thus in principle it determines the loading capacity. The last equation represents the mass conservation. By solving this system of equations with the initial condition $C_b(0) = C_a(0) = 0$ (no fines initially bonded to the carrier and no aggergates initially present in the LH300 aliquot), $C_f(0) = 27\%$ it is possible to obtain the variation in time of the three population concentrations. Properly choosing the model parameters the experimental results can be easily fitted, as indicated by the continuous lines in Figure 4 (a) and (b). Although this may appear just an exercise it offers a great deal of insight:

1) In order to fit the experimental trends one as to set $k_{off} \rightarrow 0$ or more generally $k_{off} \ll k_{on}$. This is in agreement with the assumption and the empirical evidence that our mixtures do not segregate, once the fines get in contact with the carrier surface they can hardly leave it.

2) In order to fit the experimental trends one as to set $k_{growth} > k_{break}$, i.e. to reach the steady state the amount of aggregated fines must be larger than the free fines.

3) Although ideally $C^{max}$ should correspond to the loading capacity this is not true in practice, as a variable amount of non-sieveble fines remains trapped inside the blend. This $C^{max}$ represents an ideal upper bound of the real loading capacity. We will demonstrate later on that this overestimation is small for a low shear blending where free fines form aggregates, $C^{max} = 25.5\%$, and much larger for high shear blending where aggregates are much smaller in size or their breakage frees fines in the blend that are not sieveble, here $C^{max} = 27\%$ like $C_f(0)$.

4) As the high shear process is much faster than the low shear one all the fitting constants are much larger for the former case (high shear: $k_{on} = 0.25$ sec⁻¹, $k_{off} = 0$ sec⁻¹, $k_{growth} = 0.6\ 1$ sec⁻¹, $k_{break} = 0.5$ sec⁻¹ and low shear: $k_{on} = 0.15$ sec⁻¹, $k_{off} = 0$ sec⁻¹, $k_{growth} = 0.15$ sec⁻¹, $k_{break} = 0.03$ sec⁻¹).

This model could in principle be predictive of mixing scale-up experiments if the scaling behaviour of the model parameters was known, however such parameters depend on the powder materials (cohesion energy in particular), on the particle size and shape distributions of the powders used, on the type of mixing process, on the fines concentration and on the scale as well. Understanding and mapping such a dependence remains for the moment a task far from being accomplished.

Finally we analysed how the size of the sieved aggregates changed with the mixing time and with the fines concentration. Figure 4 (d) shows how the aggregate size increased with mixing time, as their mass did not change owing to the reaching of the steady state (see Figure 4 panel a), this means that their number decreased. In other words as the mixing proceeded small aggregates broke freeing fines that were captured by larger aggregates keeping growing more and more, aggregates worked as attractors of non-bonded fines. The results of fixing the mixing time and looking at the aggregate size with increasing fines concentration is illustrated in Figure 4 (e). With a larger amount of non-bonded fines available the growth process of aggregates was speeded up and larger aggregates were found in the higher concentration blends. Together with the aggregate size also their polydispersion grew with the fines concentration.

To conclude we have so far demonstrated that, thanks to the presence of a steady state it is possible to provide blends whose characteristics are almost independent of the mixing time. After the onset of the steady state the only effect of increasing the mixing time is to "extract" more non-sieveble fines (aggregates or weakly bonded fines) from the blend, however the aggregate amount and poly-dispersion increases with concentration and a small amount of them whose size is comparable to the carrier one always passes the sieve.

**3.2 Loading capacity estimation**

To estimate the loading capacity itself we cannot solely relay on the sieve method, we need to combine different techniques.

The first attempt to quantify the loading capacity was done using a low shear mixing process at 23 rpm with batch size 500g; blends were prepared and sieved at nominal concentrations varying between 1 and 30% w/w. The results are shown in Figure 5 (a). According to the picture presented above the carrier capacity was saturated between 15% and 20% w/w, above that value a significant amount of aggregates appeared as well as a small and almost constant amount of free fines (roughly 0.2 %), while the amount of bonded fines did not exceed 21%. The observation of this phenomenon for carrier based DPI formulations has already been reported by other authors in literature although never quantitatively investigated [6,26].

The sieved blends were placed under the microscope to determine their actual coverage with the method above described, the results have been already presented as a typical outcome in Figure 3 (c). Consistently with the sieve analysis the coverage index grew almost linearly up to 20% concentration reaching then a plateau. The maximum value of 0.4 for the coverage index is consistent with the micrographies of Figure 3 (a) showing how, even above 20% w/w concentration, large portions of each carrier particle remained uncovered. These uncovered regions coincide with the perfectly smooth surfaces of the lactose tomahawk-shaped single crystals offering a reduced contact area to the fines. On the contrary, the micrograpies of the same figure relevant to the low concentration reveal how fines prefer to adhere to the most corrugated and irregular regions of the carrier surface, e.g. the interstitial regions between two fused tomahawk-shaped lactose single crystals. In the dynamical equilibrium established during blending, fines are continuously attaching and detaching from the carrier leaving its smooth surfaces unpopulated and preferring to sit in the intertitial regions or to coalish into pure fine aggregates rather than bonding to weakly adhesive carrier regions. Thus the maximum carrier loading capacity is not necessarily reached upon complete coverage of its surface. How large is the maximum coverage index reachable by a certain fine over a certain carrier clearly depends on fine adhesive and cohesive properties i.e. from the material properties, from the fine particle size distribution and from the ratio between fine size and carrier surface rouchness characteristic length.

The permeability measured on the sieved blends, reported in Figure 5 (b) decreased with increasing fine concentration; this is in agreement with the idea that the bonded fines represent an obstacle for the air to pass through the powder bed. Again, around 20% nominal concentration no more fine could be hosted on the carrier and all the excess fines would be in principle removed by the sieving procedure, this indeed resulted in a levelling off for the permeability as well. Notice how the error bars are much larger for small fine concentrations and vanish at large ones. This can be understood considering that the permeability $\kappa$ is inversely proportional to $\Delta p$ and even a small fluctuation of the latter will result in a large variation of the former. As the intrinsic uncertainty of the pressure drop measurement $\delta(\Delta p)$ increases as the powder is more permeable and $\Delta p \to 0$, the permeability uncertainty $\delta(\kappa)$ is expected to be magnified for low fine concentration blends.

Finally, Figure 5 (c) reports the powder density after conditioning before the permeability measurement, in this state the powder should be slightly more homogeneous than after pouring into the rheometer cuvette but still not as closely packed like after a typical tapping protocol. At small fine concentration the powder density increased as the fines simply attached to the carrier particle increasing their weight or they filled the gap between the particles, at large concentrations the fines covered partially the carrier particles increasing their cohesion energy and, like for fine cohesive powders, the density is expected to drop down again. In between a maximum must exist where the two effects compensate each other, the occurrence of a maximum is also expected from the models: the black dashed curves represent the theoretical density behaviour for a binary mixture in the closest packing hypothesis. This curves can be calculated once the bulk density of the two blend components alone are known (the coarse carrier density $\rho_c$ and the fines density $\rho_f$), i.e. starting with the larger component only and filling its interstitial regions with the fines or starting from the fine component only and filling its structural voids with the larger particles [27]

$$\rho(C_f) = \begin{cases} \dfrac{\rho_c}{100-C_f}100 & 0 \leq C_f \leq C_{max} \\ \dfrac{\rho_f \rho_l}{100\,\rho_f + C_f(\rho_l - \rho_f)}100 & 1 \geq C_f \geq C_{max} \end{cases} \qquad (5)$$

here $C_f$ is the % w/w of fine concentration and $\rho_l$ is the lactose true density and $C_{max} = \rho_c/(\rho_c + \rho_f - \rho_c\rho_f/\rho_l)$ represents the maximum packing density. The two curves represent an upper bound for the density of real powder mixtures, a deviation from this ideal behaviour is known to occur (also shifting the maximum position) due to poly-dispersion and cohesion of real powders which limit the possibility to reach the closest packing. Although the nominal concentration increased above 20%, the sieved blends did not contain more fines as the excess with respect to the maximum loading capacity was sieved out, thus, instead of decreasing further, the density levelled off.

The aerosolization performance was assessed with capsules containing 20 mg of the sieved blends at different nominal concentrations. The results for fine particle mass and fine particle fraction are shown in Figure 5 (d).

As previously stated, the aerosolization performance was first of all investigated to assess the uniformity of distribution of fines. The variability observed in the fine particle fraction as well as in fine particle mass was within 20% suggesting an even distribution of fines in the different blends.

As for the general behaviour of this powers upon aerosolization, the obtained data are in agreement with literature [6,7]: the emitted dose increased linearly with the loades dose in the concentration intervals investigated. It is worth nothing that the inhalation performances are quite poor simply likey because the aerodynamic diameter of LactoHale® 300 was slightly lhigher than 5μm.

Differently from the cited literature where active pharmaceutical ingredients were use, here the carrier/fine blends were comoposed by the same material, i.e. lactose of different PSD. Obviously this makes a significant difference in terms of forces involved in the aerosolization process: only cohesive phenomena could be observed without any adhesive contribution. Nevertheless, it is important to underline that similarly to what reported in Young and coworkes papers [6–8] a fine concentration threshold can be observed in relation with the FPF and FPD behaviour. Here below 20% fines the FPF decreased and the FPD remained almost constant, while above 20% the FPF levelled off and an increase, though not statistically significant, of FPD was observed. Interestigly, Young and coworkers (2005) reported a practically superimposable behaviour with lactose-salbutamol sulphate blends in a API concentration range up to 1% w/w, the opposite being observed with the same API (2011) or beclomethasone dipropionate (2018) in blends with fine concentration up to 20 and 30% respectively.

This data with those obtained with the other techniques presented here coherently point at 20% w/w as the maximum loading capacity of LactoHale® 300 over Capsulac® 60.

**3.3 Loading capacity and blending process**

So far we have discussed how to estimate the loading capacity of a given pair of carrier and fine powders by preparing different blends at increasing fine concentration, the low-shear blending process remaining identical for all of them. Therafter we have repeated the same operation for different blending processes or process parameters. The idea was to verify if and to which extent the loading capacity could be controlled by the mixing process instead of re-engineering the carrier or, more generally, by the formulation design. The rationale beyond such approach is two fold and mainly of practical nature: from one side the use of commercially available carriers "out of the box" is certainly much cheaper than modifying them or engineering a new carrier from scratch, the advantage of tailoring the loading capacity by acting on the blending process parameters rather than on the carrier is thus evident; some times the need of fine-tuning the loading capacity arises in the late stages of the drug product development when the formulation design should be already concluded and a change of excipient or the introduction of a new process step could cause large delays and regulatory complications.

According to the literature on simple mixing systems [28] the powder behaviour in a tumbling mixer drum results from the competition and the balancing/unbalancing of three main different contributions: the centrifugal force; friction forces (wall-powder friction and powder internal friction) and the gravitational force. Depending on which one is dominating three main categories of motion can be defined: slipping, the powder mass moves like a unique rigid body with respect to the drum walls, no internal motion occurs within the powder, here friction is the main player; tumbling, centrifugal force balance gravity, shear forces between powder layers occur, the powder particles move in closed loops so that no net transport occurs through the overall powder volume preserving the steady state motion; cataracting, centrifugal force dominate, small powder volumes detach from the main mass and fly across the drum to fall back on the powder free surface, the overall powder volume and shape changes continuously and randomly. With our mixing conditions we certainly lay in the tumbling regime or at the border between tumbling and cataracting regimes. In this situation the state of motion can be described by two characteristic dimensionless numbers, the Froude number $Fr$, i.e. the ratio between centrifugal and gravitational forces, and the drum filling index $\ell$:

$$Fr = \frac{\omega^2 R}{g} \quad (6)$$

$$\ell = \frac{l}{h} \quad (7)$$

$\omega$ being the angular velocity of the drum (in rad/sec), $g$ the gravitational acceleration, $R$ and $h$ the radius and height of the drum approximated to a cylinder, and $l$ the filling height of the powder within the drum. By increasing the mixing speed from 23 to 72 rpm we rose the Frounde number by an order of magnitude from $Fr \cong 0.04$ to $Fr \cong 0.4$. On the other hand, by reducing the batch size from 500g to 250g we lowered the filling index from $\ell \cong 0.34$ to $\ell \cong 0.18$. In both cases this results promoted a transition between a tumbling motion to a cataracting one [28] probably enabling direct collisions between powder blocks and drum walls, or at least in strengthening the tumbling motion with a more energetic and deeper cascading layer and a reduced volume for the plug flow region. In both cases the powder was expected to be subject to larger mechanical stress either normal stresses, due to collisions with walls, and shear stresses, in the cascading layer, with respect to the previously analysed case of 23 rpm and 500g batch size. A comparison among the previous mixing process and the two new cases of 72 rpm, 500g batch size and 72 rpm, 250g batch size is given in Figure 6, the total mixing time was kept at 3 hours. The amount of sieved aggregates was remarkably reduced by increasing the mixing speed and a further reduction was observed lowering the batch size and thus the filling height. A limited amount of aggregates does not mean necessarily that more fines were bonded to the carrier, indeed the amount of sieved free fines was larger for the 72 rpm cases compared to the 23 rpm case, suggesting that the enhanced mechanical stress felt by the powder led to aggregate breakage with consequent dispersion of fines in the blend. The coverage index, reported in panel (c) of Figure 6, was not significantly modified by the change of the process parameters confirming that a larger amount of free non-sieveble fines remained trapped inside the blend. The final proof of this can be found in the permeability measurements plotted in panel (d) of Figure 6: the lower permeability found in the region 15-27% was the fingerprint of a larger content of fines in the blend. We therefore concluded that the modification of the process parameters of a low shear mixing process was not sufficient to induce significant modifications of the carrier loading capacity.

To complete the section we compared the low shear blending results with middle shear and high shear blending processes. Figure 7 (a) and (b) show the comparison among the 23 rpm low shear mixing process at 500g batch size, the equivalent tumbling mixing performed using the middle shear with both 500g and 250g batch size, the high shear mixing process at 250 rpm for 5 minutes on 1 kg batch size. Moving from low, to middle, to high shear mixing the total mass of aggregated fines decreased considerably, due to the increase in the mechanical stress of the blends, however no significant improvement of the carrier coverage occurred, (Figure 7, panel c). Also for the middle shear blending, the reduction of the filling index resulted in strengthening the mechanical stress with a further reduction of aggregates. At low concentrations both the middle and high shear processes gave rise to a lower coverage compared to the low shear process; increasing the concentration the high shear blend coverage afforded higher values, comparable to the low shear ones, while the middle shear process maintained a lower coverage index. When repeating the blending with the middle shear drum at half filling index the coverage returned almost the same as for the low shear reference. This findings suggest that increasing the number of shear planes in the powder during blending might lead to a better dispersion of the fines but at the same time promotes the shearing motion between carrier particles favouring the fine detachment and resulting, on average, in a lower carrier coverage. As the filling index is lowered the powder motion inside the drum has a transition from cascading to cataracting and the drum geometry is no longer able to promote the increase in the shear planes amount, as a consequence the blending procedure resembles closely the low shear one. In the high shear mixer the increase of fine concentration seems to lower the effect of the shear stresses leading to an increase of carrier coverage, this is in agreement with the recent findings of Hertel et al. [29] who proposed a model equation to account for this:

$$P = \frac{\rho\,(E - E_0)}{m} \qquad (8)$$

Where $P$ is the pressure/stress acting on the powder obtained as the energy imparted by the blade on the blend $E$ divided by the total powder volume, i.e. the powder mass $m$ divided by its bulk density $\rho$ ($E$ is obtained from the power supplied to the blade engine multiplied by the mixing duration. To cut out the energy losses due to friction and mechanical vibrations of the mixer, $E$ is substracted by $E_0$ which is energy consumption measured in a run with the empty bin). From equation (8) it is clear that increasing the fine concentration, thus reducing the blend bulk density the pressure/stress exerted on the powder decreases and a carrier coverage similar to that obtained with a low shear mixing process can in principle be found.

Based on the comparison of the fine aggregate mass only some of the different mixing processes might be considered as equivalent: comparing Figure 6 (a)-(b) and Figure 7 (a)-(b) the sieved aggregate curve for the middle shear 23 rpm, 250g almost perfectly overlapped with that for the low shear 72rpm, 500g; the same applied for the high shear curve and the middle shear 23 rpm, 500g. The carrier coverage of these process was however not the same, thus the fine aggregate mass alone was not a good indicator of the equivalence of two mixing processes.

Finally, panel (d) of Figure 7 reports the permeability measurements as a consistency check. The middle shear process at 23 rpm, 250g batch size showed a coherent behaviour, it had a smaller mass of aggregates compared to the reference low shear process but almost the same coverage index, thus we expected to find more non-sieveble fines inside the blends, indeed the permeability in this case was lower than in the reference low shear process. However, for the middle shear process at 23 rpm, 500g as well as for the high shear one, the permeability seemed to behave incoherently. For the middle shear process the permeability was perfectly overlapped with the low shear case despite it had less aggregate fines and a lower carrier coverage. For the high shear process at lower concentration the permeability was even higher than the low shear reference. This apparent inconsistency stems from the fact that the permeability of a blend was determined by the competition of two effects: from one hand, the presence of free fines is expected to lower the permeability as they clog the air percolation paths; on the other hand a lower carrier coverage increases the number and the average size of the percolation channels favouring the air passage and thus increasing the permeability. For this reason the permeability can be used to compare the amount of non-bonded fines in two blends only if the relevant carrier coverage is the same. Both for the middle shear process at 23 rpm, 500g and the high shear one at low fine concentration, the low coverage index contribution, that would increase the permeability, dominanted with respect to the free fines contribution, that would reduce the permeability. For the high shear process, when increasing the concentration the carrier coverage became identical to the low shear one, the permeability lowered significantly with respect to the low shear case due to the sole contribution of the more abundant non-sieved fines.

The aerosolization performances of the powder mixture obtained with mid and high shear process (Figure 8) practically overlapped to those of the blend obtained with low shear process presented in Figure 5 (d). Non significant difference could be observe among the three mixing processes. This is somehow unexpectcted especially in the light of the results presented in Figure 6 (a)-(b) and 7 (a)-(b). To explain this apparently peculiar behavior it should be borne in mind once again that the fines that give riese to the sievable aggregates were identical to the carried in terms of chemical composition, thus, excluding the effect of the exposed area, the "cohesive" forces which kept togheter the non sievable fine, responsible in proinciple of a poor aerosolization, here coincide with the adhesive forces with the coarse carrier. As reported by Thalberg at al. [30] fine lactose has higher dispersibility than a API such as beclomethasone dipropionate. Therefore, it is not surprising that a high shear process that, generates higher amount of free fines, did not afford better aerosolisation, as the capability of the aggregates to separate into sigle small unit during the aerosolization was comparable to that of the fines to detach from the carrier.

## 4. Conclusions

In this paper we proposed and discuss a method to estimate the loading capacity of carrier particles, a key characteristic in the design of DPI carrie based formulations. We also examined the behaviour of the blends when the loading capacity is saturated and no more fines can be allocated on the carrier surface. The method is robust, reproducible and fast and based on simple equipment often found in powder rheology laboratories. It combines three different measurement techniques none of which, considered alone, could provide complete or unambiguous results, however, they complement each other giving a coherent quantitative picture of a carrier loading capacity.

As a test case we focussed on a coarse lactose carrier hosting lactose fines intended as a surrogate of a real API. The qualitative behaviour of real API fines compared to lactose ones is the same both below and above the loading capacity saturation. The loading capacity itself, the maximum coverage index reachable and the propension of the non-bonded fines to form aggregates, depend clearly on the specific couple of API and carrier chosen, more precisely from their material properties, their roughness and size. Evidences that beclomethasone dipropionate have a similar qualitative behaviour of LH300 upon saturating the loading capacity can be found in the work of Yeung et al. [8], for salbutamol sulfate a good reference paper is the one by Hertel et al. [26].

In the second part of this work we studied the impact of the mixing process conditions and parameters on the carrier loading capacity employing the newly introduced characterization method. For our simplified model blends, the loading capacity is only marginally dependent on the mixing conditions and the only way to drastically modify it is through a formulation change, e.g. by changing the surface characteristics, shape and morphology of the carrier particles.

The aerosolization performances are dominated by the adhesive forces that are practically equal between the couples carrier-fine and fine-fine, being the two components of the blend similar in nature. The main general conclusion that can be drawn from the aerosol performance data presented here in comparison with the literature data is that the boundary conditions becomes less relevant when the adhesive/cohesive forces ratio tends to the unit value.

The method presented in this paper can certainly be improved. The main problem remains the estimation of the amount of unsieved fines present in the blend once the carrier loading capacity is saturated. The permeability measurement offers indeed an indirect evidence of the presence of unsieved fines in the blends however, it is not a quantitative technique and it does not not provide information about the state of these fines. Other rheological properties of the blends might be more sensitive to the presence of the unsieved fines and might give quantitative results after implementation of a calibration procedure.

The proposed loading capacity estimation technique will be tested for various carrier and API having different size and morphology characteristics to challenge its reliability in those cases where the result is predictable or expected on the basis of simple geometrical considerations.

## Figures and Captions

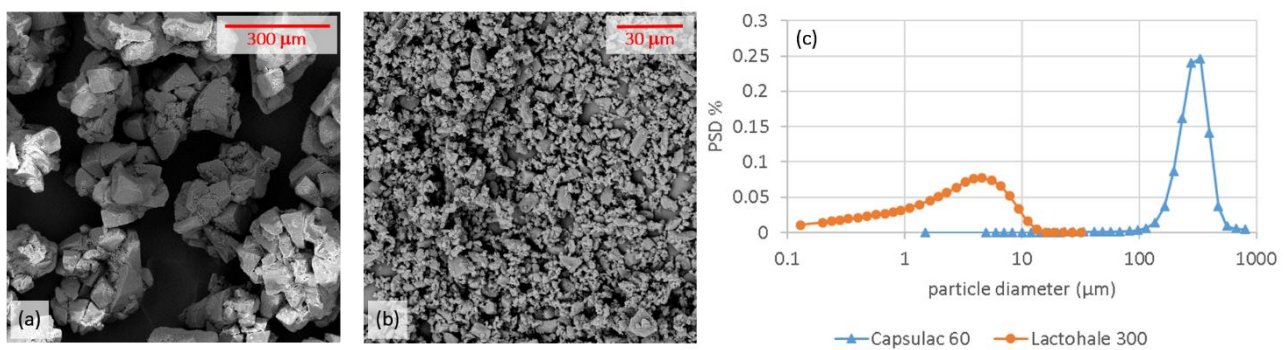

**Figure 1**: SEM images of CapsuLac® 60 (Meggle) and LactoHale® 300® (DFE Pharma), panel (a) and (b) respectively. (c) PSD distribution of the two excipient obtained with a Helos BR from Sympatec® using 0.5 bar with an R5 lens for Capsulac® 60 and 2.0 bar with an R1 lens for Lactohale® 300.

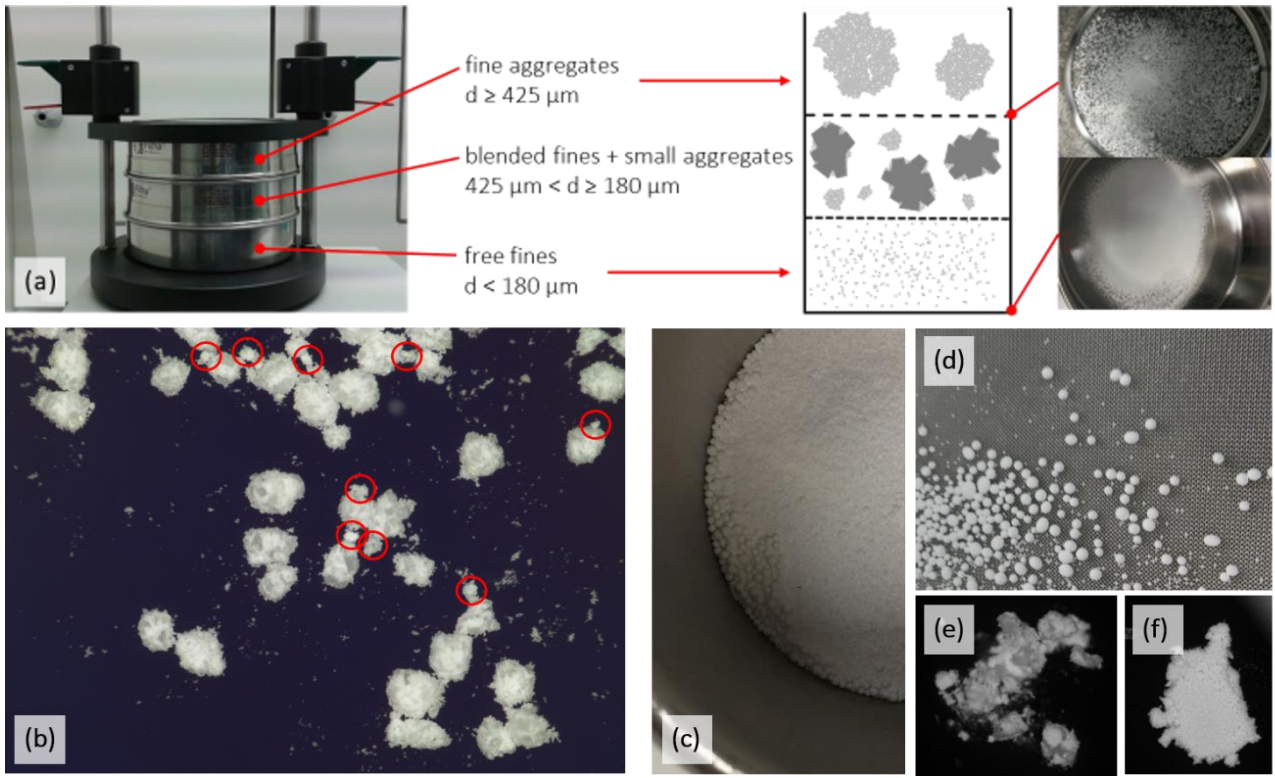

**Figure 2**: (a) Setup of the sieve shaker with two sieves and a collection bin, the right part of the panel shows a pictorial representation of sieves and collection bin content together with two pictures of the sieves after the analysis of a 30% w/w blend. (b) micrography of 30% w/w blend collected in the 180 μm sieve, some of the small fine aggregates that cannot be separated from the blend are highlighted by the red circles. (c) and (d) close-up pictures of aggregates in the mixing bin and on the sieve net. (e) and (f) are micrographies of a blend particle and an aggregate directly broken under the microscope, in the first case fragments of the Capsulac® 60 particle are clearly visible, in the latter case nothing except LactoHale® 300 fines are visible "inside" the aggregate.

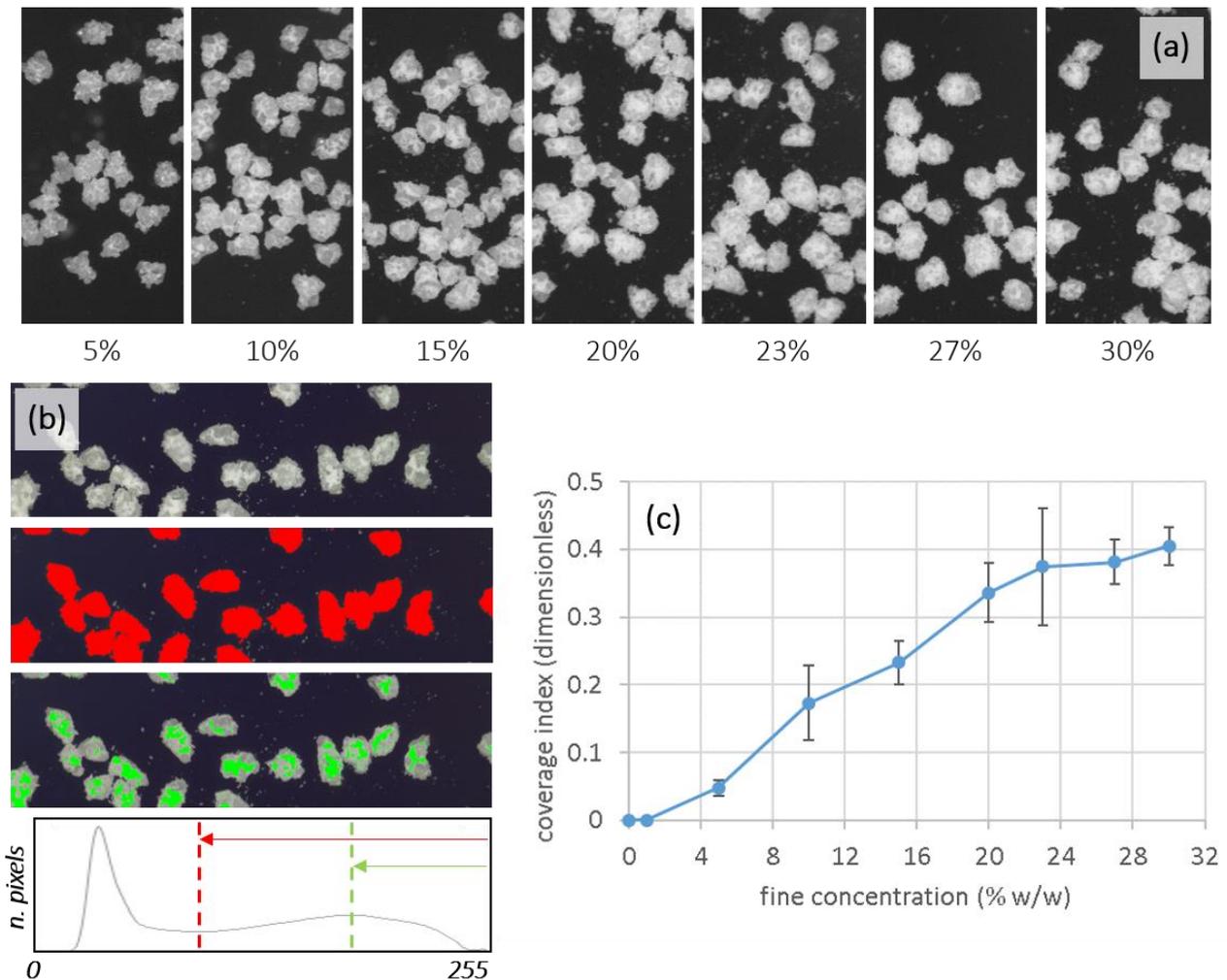

**Figure 3**: (a) micrographies of low shear blends at increasing fine concentration after sieving analysis. (b) Example of image filtering on the image brightness to localize the fine patches on the blended carrier particles and measure the particle (red) and patches (green) area. The bottom panel shows the brightness histogram of the image, i.e. the number of bixels pertaining to each brightness number from 0 to 255, the red and green dashed lines represent the thresholds used to map the full blend particle and the fine patches on it. (c) plot of the coverage index as a function of the fine concentration calculated on the blends of panel (a), the amplitude of the vertical bars represent the data standard deviation.

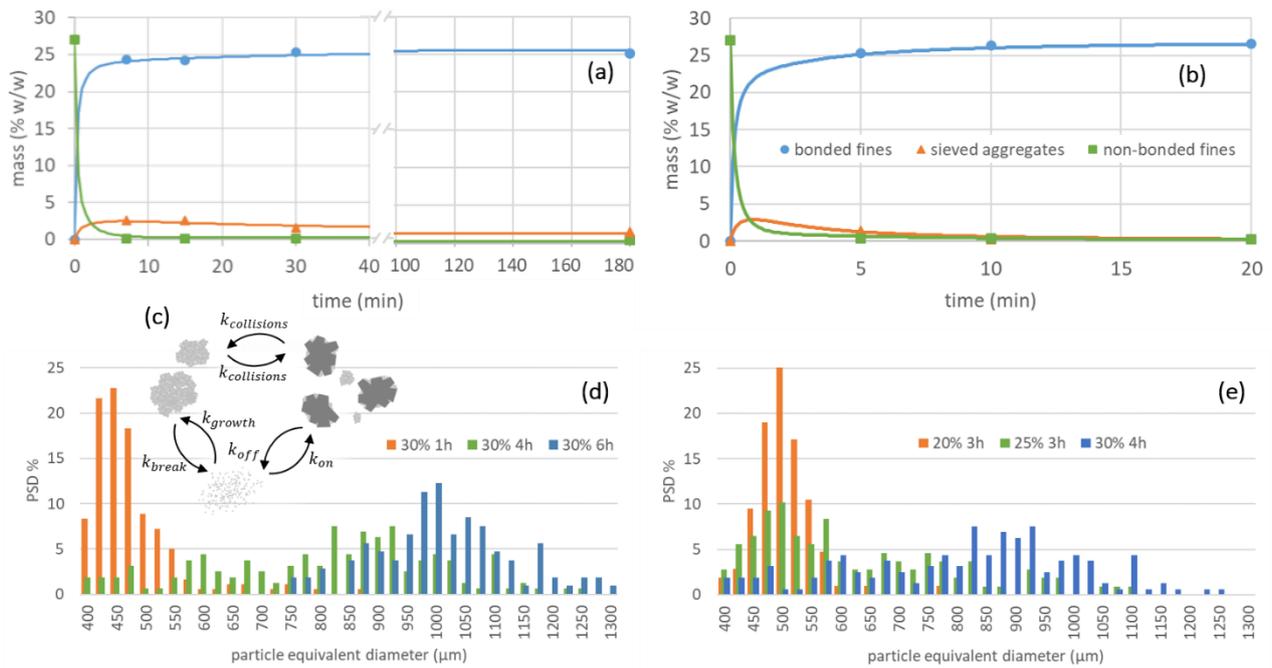

**Figure 4**: (a) mass of sieved fractions as a function of mixing time for a low shear process of 250g blend at 27% w/w concentration, 72 rpm speed. Dots are experimental data, continuous lines come from a first order reaction model with parameters described in the text. (b) same as (a) but for a high shear process applied to 1Kg of blend at 27% w/w concentration at 250 rpm speed. (c) sketch of the dynamic equilibrium between the three populations of free, bonded and agglomerated fines, the different constants labelled by the k letters are the reaction rates for the different reaction channels. For the low shear mixing process panels (d) and (e) display the histograms of the equivalent diameter of sieved aggregates measured through image analysis of the optical micrographies, the first plot shows the results for the same blend concentration, 30% w/w, varying the mixing time while the second one shows the effect of an increasing concentration.

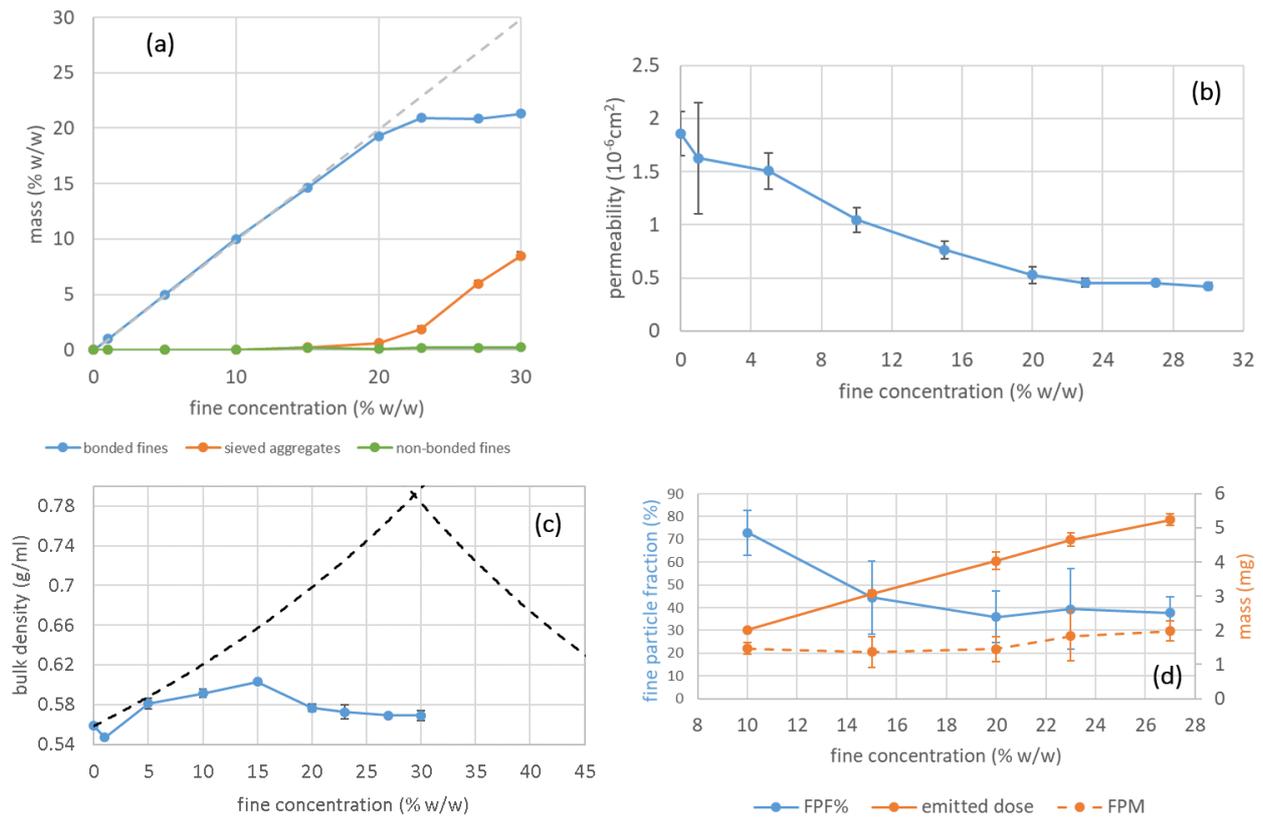

**Figure 5:** (a) moss of the free, bonded and aggregated fines as a function of the fines concentration for the low shear micing process at 500g batch size. The dashed line represents the bisecting line of the plot where the measure % mass correspond to the nominal % fine concentration. As long as the blue line is superimposed to the dashed one all the fines are bonded to the carrier, when it departs from it non-bonded fines appear. (b) permeability of the sieved blends as a function of fines concentration. (c) density of the sieved blends after conditioning cycle in the FT4 rheometer as a function of the fines concentration. The dashed lines repreent the calculated theoretical densities in the closest packing hypotheisis for our binary mistures according to equation (5). (d) inhalation performances as a function of fines concentration in terms of fine particle fraction (FPF) normalized to the emitted dose, emitted dose and fine particle mass (FPM).

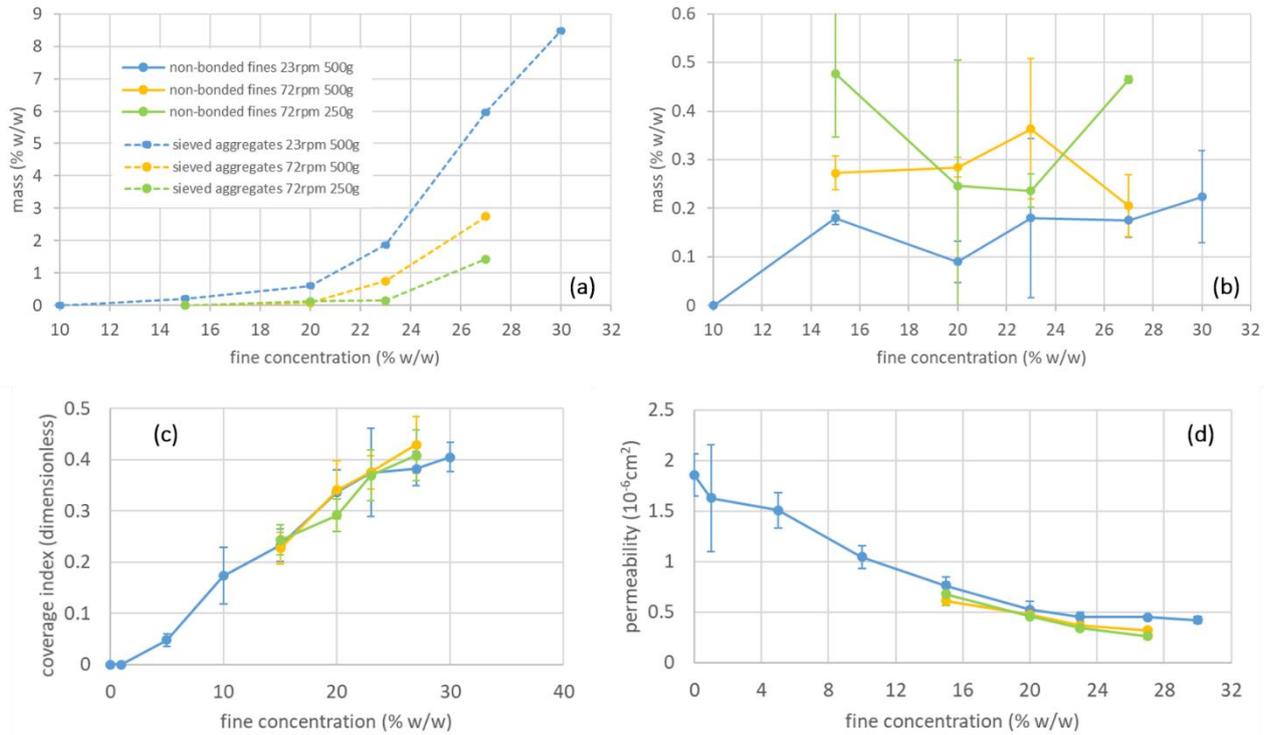

**Figure 6**: mass of the aggregated (a) and free fines (b) as a function of the fines concentration for the low shear mixing process at 72 rpm with 500 and 250g batch size, the results for the 23 rpm, 500g of Figure 5 are also presented for comparison. (c) corresponding coverage index as a function of the fines concentration. (d) corresponding permeability of the sieved blends as a function of fines concentration.

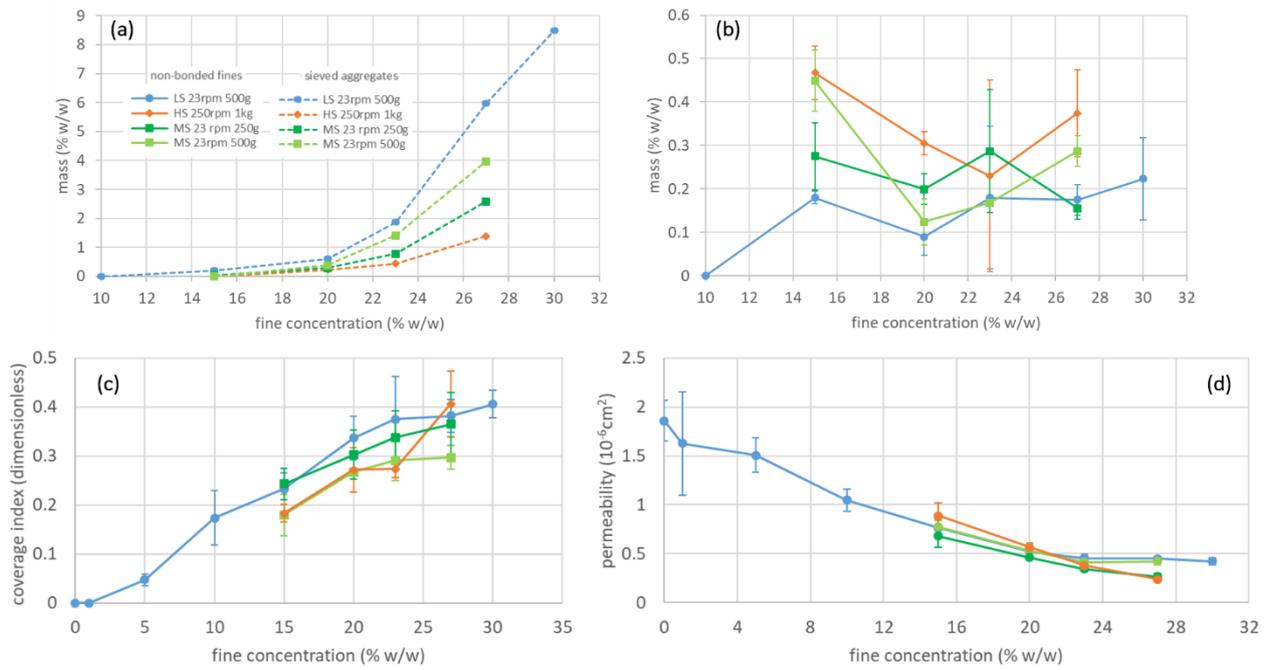

**Figure 7**: mass of the aggregated (a) and free fines (b) as a function of the fines concentration for the low shear mixing process at 23 rpm, 500g batch size, middle shear process at 23 rpm, 500 and 250g batch size and high shear process 250 rpm, 1Kg batch size. (c) corresponding coverage index as a function of the fines concentration. (d) corresponding permeability of the sieved blends as a function of fines concentration.

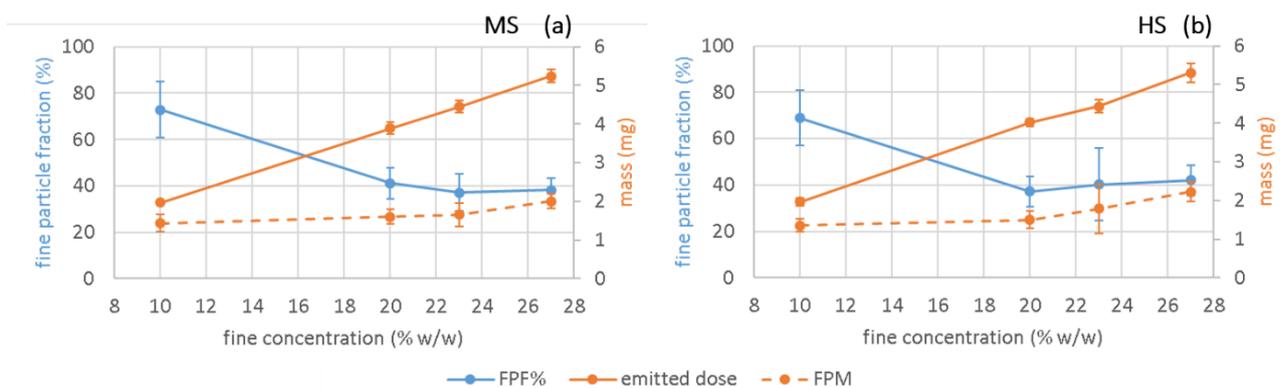

**Figure 8**: (a) and (b) Inhalation performances of the middle shear (MS) and high shear (HS) blending as a function of fine concentration in terms of fine particle fraction (FPF) normalized to the emitted dose, emitted dose and fine particle mass.

# Bibliography

[1]   A.H. De Boer, H.K. Chan, R. Price, A critical view on lactose-based drug formulation and device studies for dry powder inhalation: Which are relevant and what interactions to expect?, Adv. Drug Deliv. Rev. 64 (2012) 257–274. doi:10.1016/j.addr.2011.04.004.

[2]   C. Pitchayajittipong, R. Price, J. Shur, J.S. Kaerger, S. Edge, Characterisation and functionality of inhalation anhydrous lactose, Int. J. Pharm. 390 (2010) 134–141. doi:10.1016/j.ijpharm.2010.01.028.

[3]   A.H. De Boer, D. Gjaltema, P. Hagedoorn, H.W. Frijlink, Can "extrafine" dry powder aerosols improve lung deposition?, Eur. J. Pharm. Biopharm. 96 (2015) 143–151. doi:10.1016/j.ejpb.2015.07.016.

[4]   A. Della Bella, E. Salomi, F. Buttini, R. Bettini, The role of the solid state and physical properties of the carrier in adhesive mixtures for lung delivery, Expert Opin. Drug Deliv. 15 (2018) 665–674. doi:10.1080/17425247.2017.1371132.

[5]   W. Kaialy, A review of factors affecting electrostatic charging of pharmaceuticals and adhesive mixtures for inhalation, Int. J. Pharm. 503 (2016) 262–276. doi:10.1016/j.ijpharm.2016.01.076.

[6]   P.M. Young, O. Wood, J. Ooi, D. Traini, The influence of drug loading on formulation structure and aerosol performance in carrier based dry powder inhalers, Int. J. Pharm. 416 (2011) 129–135. doi:10.1016/j.ijpharm.2011.06.020.

[7]   P.M. Young, S. Edge, D. Traini, M.D. Jones, R. Price, D. El-Sabawi, C. Urry, C. Smith, The influence of dose on the performance of dry powder inhalation systems, Int. J. Pharm. 296 (2005) 26–33. doi:10.1016/j.ijpharm.2005.02.004.

[8]   P.M. Young, A. Tweedie, D. Lewis, T. Church, D. Traini, Limitations of high dose carrier based formulations, Int. J. Pharm. 544 (2018) 141–152. doi:10.1016/j.ijpharm.2018.04.012.

[9]   F. Grasmeijer, P. Hagedoorn, H.W. Frijlink, A.H. de Boer, Mixing Time Effects on the Dispersion Performance of Adhesive Mixtures for Inhalation, PLoS One. 8 (2013) 1–18. doi:10.1371/journal.pone.0071339.

[10]  K. Thalberg, E. Berg, M. Fransson, Modeling dispersion of dry powders for inhalation. the concepts of total fines, cohesive energy and interaction parameters, Int. J. Pharm. 427 (2012) 224–233. doi:10.1016/j.ijpharm.2012.02.009.

[11]  F. Grasmeijer, A.J. Lexmond, M. Van Den Noort, P. Hagedoorn, A.J. Hickey, H.W. Frijlink, A.H. De Boer, New mechanisms to explain the effects of added lactose fines on the dispersion performance of adhesive mixtures for inhalation, PLoS One. 9 (2014) 1–11. doi:10.1371/journal.pone.0087825.

[12]  W. Kaialy, On the effects of blending, physicochemical properties, and their interactions on the performance of carrier-based dry powders for inhalation-A review, Adv. Colloid Interface Sci. 235 (2016) 70–89. doi:10.1016/j.cis.2016.05.014.

[13]  J. Rudén, A. Vajdi, G. Frenning, T. Bramer, Comparison Between High and Low Shear Mixers in the Formation of Adhesive Mixtures for Dry Powder Inhalers, Respir. Drug Deliv. Eur.

2019. (2019) 457–460.

[14] G. Pilcer, N. Wauthoz, K. Amighi, Lactose characteristics and the generation of the aerosol, Adv. Drug Deliv. Rev. 64 (2012) 233–256. doi:10.1016/j.addr.2011.05.003.

[15] J. Rudén, G. Frenning, T. Bramer, K. Thalberg, G. Alderborn, Relationships between surface coverage ratio and powder mechanics of binary adhesive mixtures for dry powder inhalers, Int. J. Pharm. 541 (2018) 143–156. doi:10.1016/j.ijpharm.2018.02.017.

[16] A. Vasilenko, B.J. Glasser, F.J. Muzzio, Shear and flow behavior of pharmaceutical blends - Method comparison study, Powder Technol. 208 (2011) 628–636. doi:10.1016/j.powtec.2010.12.031.

[17] C.C. Sun, Quantifying effects of moisture content on flow properties of microcrystalline cellulose using a ring shear tester, Powder Technol. 289 (2016) 104–108. doi:10.1016/j.powtec.2015.11.044.

[18] S. Koynov, B. Glasser, F. Muzzio, Comparison of three rotational shear cell testers: Powder flowability and bulk density, Powder Technol. 283 (2015) 103–112. doi:10.1016/j.powtec.2015.04.027.

[19] Y. Nakayama, R.F. Boucher, Introduction to Fluid Mechanics, Butterworth Heinemann, Oxford, 1999.

[20] X. Fu, D. Huck, L. Makein, B. Armstrong, U. Willen, T. Freeman, Effect of particle shape and size on flow properties of lactose powders, Particuology. 10 (2012) 203–208. doi:10.1016/j.partic.2011.11.003.

[21] E. Cordts, H. Steckel, Capabilities and limitations of using powder rheology and permeability to predict dry powder inhaler performance, Eur. J. Pharm. Biopharm. 82 (2012) 417–423. doi:10.1016/j.ejpb.2012.07.018.

[22] R. Holdich, Fundamentals of Particle Technology, Midland Information Technology and Publishing, Shepshed, Leicestershire U.K., n.d.

[23] R. Freeman, Measuring the flow properties of consolidated, conditioned and aerated powders - A comparative study using a powder rheometer and a rotational shear cell, Powder Technol. 174 (2007) 25–33. doi:10.1016/j.powtec.2006.10.016.

[24] D. Geromichalos, M.M. Kohonen, F. Mugele, S. Herminghaus, Mixing and Condensation in a Wet Granular Medium, Phys. Rev. Lett. 90 (2003) 4. doi:10.1103/PhysRevLett.90.168702.

[25] F. Podczeck, Particle-Particle Adhesion in Pharmaceutical Powder Handling, Imperial College Press, London, 1997.

[26] M. Hertel, E. Schwarz, M. Kobler, S. Hauptstein, H. Steckel, R. Scherließ, Powder flow analysis: A simple method to indicate the ideal amount of lactose fines in dry powder inhaler formulations, Int. J. Pharm. 535 (2018) 59–67. doi:10.1016/j.ijpharm.2017.10.052.

[27] R.L. Brown, J.C. Richards, Principles of Powder Mechanics, Pergamon Press, Oxford, 1970.

[28] J. Mellmann, The transverse motion of solids in rotating cylinders-forms of motion and transition behavior, Powder Technol. 118 (2001) 251–270. doi:10.1016/S0032-5910(00)00402-2.

[29]  M. Hertel, E. Schwarz, M. Kobler, S. Hauptstein, H. Steckel, R. Scherließ, The influence of high shear mixing on ternary dry powder inhaler formulations, Int. J. Pharm. 534 (2017) 242–250. doi:10.1016/j.ijpharm.2017.10.033.

[30]  K. Thalberg, S. Åslund, M. Skogevall, P. Andersson, Dispersibility of lactose fines as compared to API in dry powders for inhalation, Int. J. Pharm. 504 (2016) 27–38. doi:10.1016/j.ijpharm.2016.03.004.